\documentclass
[aps,preprint,preprintnumbers,amsmath,amssymb,nofootinbib]{revtex4}%
\usepackage{amssymb}
\usepackage{amsmath}
\usepackage{bm}
\usepackage{amsthm}
\usepackage{graphicx}
\usepackage{amsfonts}%
\providecommand{\U}[1]{\protect\rule{.1in}{.1in}}
\theoremstyle{plain}

\theoremstyle{definition}

\theoremstyle{proposition}

\theoremstyle{lemma}

\theoremstyle{corollary}

\begin{document}
\title{An Invariant Approach to Weyl's unified field theory }

\begin{abstract}
We revisit Weyl's unified field theory, which arose in 1918, shortly after
general relativity was discovered. As is well known, in order to extend the
program of geometrization of physics started by Einstein to include the
electromagnetic field, H. Weyl developed a new geometry which constitutes a
kind of generalization of Riemannian geometry. However, despite its
mathematical elegance and beauty, a serious objection was made by Einstein,
who considered Weyl's theory not suitable as a physical theory since it seemed
to lead to the prediction of a not yet observed effect, the so-called "second
clock effect\textquotedblright\ . In this paper, our aim is to discuss Weyl's
proposal anew and examine its consistency and completeness as a physical
theory. Finally, we propose new directions and possible conceptual changes in
the original work. As an application, we solve the field equations assuming a
Friedmann-Robertson-Walker universe and a perfect fluid as its source.
Although we have entirely abandoned Weyl's atempt to identify the vector
field with the $4$-dimensional electromagnetic potentials, \ which here must
be simply viewed as part of the space-time geometry, we believe that in this
way we could\ perhaps be led to a rich and interesting new modified gravity
theory. 

\end{abstract}
\author{$^{1}$T. A. T. Sanomiya, $^{2}$I. P.\ Lobo, $^{1}$J. B. Formiga,\ $^{1}$F.
Dahia and $^{1}$C. Romero}
\affiliation{$^{1}$Departamento de F\'{\i}sica, Universidade Federal da Para\'{\i}ba, Caixa
Postal 5008, 58059-970, Jo\~{a}o Pessoa, PB, Brazil.}
\affiliation{$^{2}$Departamento de F\'{\i}sica, Universidade Federal de Lavras, Caixa
Postal 3037, 37200-000, Lavras, MG, Brazil}
\affiliation{E-mail: sanomiya@fisica.ufpb.br; iarley\_lobo@fisica.ufpb.br;
jformiga@fisica.ufpb.br, fdahia@fisica.ufpb.br,\ cromero@fisica.ufpb.br.\ }
\maketitle

\section{Introduction}

In his attempt to unify gravity with electromagnetism H. Weyl discovered a new
geometry, which in a certain way, constitutes a kind of generalization of
Riemannian geometry \cite{Weyl}. As he wrote in the introduction of his
original paper, the insight which led him to a new geometry came from the
perception that the Riemann theory of parallel transport was entirely
dependent on concepts directly taken from our intuition of the "rigid"
Euclidean spaces . Indeed, in this geometric setting the parallel transport of
a vector $V$\ along a certain path is required to preserve the length of $V$.
In more technical terms, this requirement imposed on a manifold endowed with a
metric tensor $g$ and a connection $\nabla$ arises as a direct consequence of
what is known in the literature as the compatibility condition between $g$ and
$\nabla$. Then, from Koszul formula, it follows the celebrated Levi-Civita
theorem, which states that for torsion-free manifolds there exists a unique
connection completely determined by the metric \cite{Manfredo}. Weyl, however,
found the Riemannian compatibility condition too restrictive, and replaced it
by a much weaker form, which then allows for the variation of the length of
vectors along parallel transport, this process being regulated by a new
geometric object, namely a 1-form field $\sigma$, later to be identified with
the electromagnetic four-potential\textit{. }The introduction of the 1-form
field $\sigma$ leads, in turn, to a new notion of curvature, a sort of "length
curvature" (\textit{Streckenkrummung})\ \ \ in addition to the "direction
curvature" (\textit{Richtungkrummung}), the latter represented by the Riemann
tensor. The length curvature is quantified by the 2-form $F=d\sigma$, whose
mathematical properties present striking similarities with those possessed by
the electromagnetic tensor. After this first development, another important
discovery made by Weyl was that his geometric construction exhibited a new
kind of symmetry. Indeed, he found that his modified compatibiliy condition,
as well as the lengh curvature, were both invariant under a certain group of
transformations involving $g$ and $\nabla$. It is worth mentioning that the
discovery of this new symmetry, \ later to be called \textit{gauge symmetry}
(in addition to the already known general relativistic invariance under
space-time diffeomorphisms) is now viewed as a most significant fact in the
history of physics: it represents the birth of modern gauge theories
\cite{Raifertaigh}. It turned out that \textit{ Weyl's Principle of Gauge
Invariance} played an essential role in the development of the unified field
theory. Indeed, in building the action for the gravitational and
the\ (geometric) electromagnetic fields, Weyl was primarely guided by this
principle and chose the simplest of all possible invariants. He also took
advantage of the principle to work out the field equations in a particular
gauge (the "natural" gauge), where the field equations look much more simple.
Enriched by the property of gauge symmetry, the geometric structure of
space-time in Weyl's theory became more complex, rather similar to what is
known as a conformal structure, that is, a manifold equipped with an
equivalence class of triples $\mathcal{M=}\left\{  (g,\nabla,\sigma)\right\}
$, \ in \ which the members of the class are related by Weyl transformations
and satisfy a particular compatibility condition. It is to be expected that in
such space-time only invariants (in the sense of Weyl's principle) may have
physical meaning. (For instance, as Weyl put it clearly, the usual metric
concept of length is no longer meaningful \cite{Weyl}.) This entirely new
framework has far-reaching consequences as far as\ as it selects which
physical scenarios are allowed to come in, and it is our aim in the present
work to investigate some of these possibilities under the guide of gauge
invariance hoping in this way to carry on with Weyl's original program.\ With
regard to the latter point we think two questions must be addressed. First, to
what extent did Weyl succeed in constructing a unifying theory of gravity and
electromagnetism? Second, is the theory free from inconsistency and/or
incompleteness? \textit{ }(For the reader interested in historical and
philosophical issues concerning Einstein's critical review of Weyl's unified
theory see, for instance, \cite{Goenner}, and references therein).

The paper is organized as follows. In Section 2, we give a brief summary of
Weyl geometry. We then proceed to Section 3 to present the Weyl field
equations, both written in an arbitrary and in the\textit{ natural gauge}. In
Section 4, we discuss the field equations in the limit when space-time becomes
Riemannian, and give an interpretation for the constant that appears in the
natural gauge. Section 5 contains a discussion of the nature of the geometric
electromagnetism introduced by Weyl and the conceptual problems arising from
this identification. Section 6 is devoted to\ the notion of time in Weyl
theory and the related problem of the second clock effect. In Section 7, we
touch on the question of how to extend Weyl theory to include matter. In
Section 8, we outline the axiomatic structure of the new approach with the aim
at defining a gauge-invariant procedure to extend the Weyl field equations to
include matter fields. As an application, in Section 9 we solve the field
equations assuming a Friedmann-Robertson-Walker universe and a perfect fluid
as its source. We conclude with some remarks in Section 10.

\section{\bigskip A brief summary of Weyl geometry}

Weyl geometry is perhaps one of the simplest generalization of Riemannian
geometry, the only modification being the fact that the covariant derivative
of the metric tensor $g$ is not zero, but instead given by \footnote{This
article was written in parallel with the authors' contribution to the
Proceedings of the 10th Alexander Friedmann Seminar on Gravitation and
Cosmology, and should be considered as a completed and streamlined version of
the latter. Therefore identical prose may be found in some parts betweenn
\ the two texts.}%

\begin{equation}
\nabla_{\alpha}g_{\beta\lambda}=\sigma_{\alpha}g_{\beta\lambda},
\label{compatibility}%
\end{equation}
where $\sigma_{\alpha}$ \ denotes the components of a one-form field
$\sigma\ $in a local coordinate basis. This weakening of the Riemannian
compatibility condition is entirely equivalent to requiring that the length of
a vector field may change when parallel-transported \ along a curve in the
manifold \cite{Pauli}. We shall refer to the triple $(M,g,\sigma)$\ consisting
of a differentiable manifold $M$ endowed with both a metric $g$ and a 1-form
field $\sigma\ $as a \textit{Weyl gauge }(or,\textit{ Weyl frame)}. Now one
important discovery made by Weyl was the following. Suppose we perform the
conformal transformation
\begin{equation}
\overline{g}=e^{f}g, \label{conformal1}%
\end{equation}
where $f$ is an arbitrary scalar function defined on $M$. Then, the Weyl
compatibility condition (\ref{compatibility}) still holds provided that we let
the Weyl field $\sigma$ transform as
\begin{equation}
\overline{\sigma}=\sigma+df. \label{gauge}%
\end{equation}
In other words, the Weyl compatibility condition does not change\ when we go
from one gauge $(M,g,\sigma)$ to another gauge $(M,\overline{g},\overline
{\sigma})$ by simultaneous transformations in $g$ and $\sigma$.

If we assume that the Weyl connection\ $\nabla$ is symmetric,\ a
straightforward algebra shows that one can express the components of the
affine connection in an arbitrary vector basis completely in terms of the
components of $g$ and $\sigma$:%
\begin{equation}
\Gamma_{\beta\lambda}^{\alpha}=\{_{\beta\lambda}^{\alpha}\}-\frac{1}%
{2}g^{\alpha\mu}[g_{\mu\beta}\sigma_{\lambda}+g_{\mu\lambda}\sigma_{\beta
}-g_{\beta\lambda}\sigma_{\mu}], \label{Weylconnection}%
\end{equation}
where $\{_{\beta\lambda}^{\alpha}\}$ represents the Christoffel symbols. It is
not difficult to see that the connection and, consequently, the geodesic
equations are invariant with respect to the transformations (\ref{conformal1})
and (\ref{gauge}).

We now present Weyl's second great discovery. Suppose we are given two vector
fields $V$ and $U$ parallel-transported along a curve $\alpha=\alpha
(\lambda).$ Then, (\ref{compatibility}) clearly leads to the following
equation:
\begin{equation}
\frac{d}{d\lambda}g(V,U)=\sigma(\frac{d}{d\lambda})g(V,U),
\label{covariantderivative}%
\end{equation}
where $\frac{d}{d\lambda}$ denotes the vector tangent to $\alpha$. If we
integrate this equation along the curve $\alpha$, starting from a point
$P_{0}=\alpha(\lambda_{0}),$ we obtain \cite{Pauli}%
\begin{equation}
g(V(\lambda),U(\lambda))=g(V(\lambda_{0}),U(\lambda_{0}))e^{\int_{\lambda_{0}%
}^{\lambda}\sigma(\frac{d}{d\rho})d\rho}. \label{integral}%
\end{equation}
Setting $U=V$ and denoting by $L(\lambda)$ the length of the vector
$V(\lambda)$ at a point\ $P=\alpha(\lambda)$\ of the curve, it is easy to
verify that in a local coordinate system $\left\{  x^{\alpha}\right\}  $ the
equation (\ref{covariantderivative}) becomes
\begin{equation}
\frac{dL}{d\lambda}=\frac{\sigma_{\alpha}}{2}\frac{dx^{\alpha}}{d\lambda}L.
\label{length}%
\end{equation}
Let us now consider the set of all closed curves $\alpha:[a,b]\in R\rightarrow
M$, i.e, with $\alpha(a)=\alpha(b).$ Then, either from (\ref{integral}) or
(\ref{length})\ it follows that
\[
L=L_{0}e^{\frac{1}{2}%
{\displaystyle\oint}
\sigma_{\alpha}dx^{\alpha}},
\]
where \ $L_{0}$ and $L$ denotes the values of $L(\lambda)$ at $a$ and $b$,
respectively. From Stokes's theorem we then can write\footnote{Here we are
assuming that the region of integration is simply connected.}
\[
L=L_{0}e^{-\frac{1}{4}\int\int F_{\mu\nu}dx^{\mu}\wedge dx^{\nu}},
\]
where $F_{\mu\nu}=\partial_{\nu}\sigma_{\mu}-\partial_{\mu}\sigma_{\nu}.$ We
thus see that, according to the rules of Weyl geometry, the necessary and
sufficient condition for a vector to have its original length preserved after
being parallel transported along any closed trajectory is that the 2-form
$F=d\sigma=\frac{1}{2}F_{\mu\nu}dx^{\nu}\wedge dx^{\mu}$ vanishes.

Therefore Weyl realized that in his new geometry there are two kinds of
curvature, a \textit{direction curvature (Richtungkrummung) }and a
\textit{length curvature (Streckenkrummung}). The first is responsible for
changes in the direction of parallel-transported vectors and is given by the
usual curvature tensor $R_{\;\beta\mu\nu}^{\alpha}$, while the other regulates
the changes in their length, and is given by $F_{\mu\nu}.$ Weyl's second great
discovery was that the 2-form $F$ is invariant under the gauge transformation
(\ref{gauge}). The analogy with the electromagnetic field is now clear and
becomes even more so when we take into account that $F$ satisfies the identity
$dF=0$ \footnote{In a local coordinate system, this identity takes the form
$\partial_{\mu}F_{\alpha\beta}+\partial_{\beta}F_{\mu\alpha}+\partial_{\alpha
}F_{\beta\mu}=0$, which looks identical to one pair of Maxwell's
equations.}$.$

\section{ The field equations of Weyl's unified field theory}

As we know, the Weyl transformations (\ref{conformal1}) and (\ref{gauge})
define a whole equivalence class in the set $\{(M,g,\sigma)\}$ of\ all Weyl
gauges. It is then natural to expect that, as in conformal geometry the
geometrical objects of interest are conformal-invariant, here we should look
for those that are gauge-invariant \footnote{In conformal geometry, one basic
invariant is the Weyl tensor $W_{\;\beta\mu\nu}^{\alpha}$. In conformal
gravity, this tensor is used to form the scalar $W_{\;\alpha\beta\mu\nu}%
W_{\;}^{\alpha\beta\mu\nu}$, which, then, defines the gravitation sector of
the action \cite{Manheim}.} Surely, these invariants will be fundamental to
build the action that is expected to give the field equations of the
geometrical unified theory. Some basic invariants are easily found: the affine
connection $\Gamma_{\beta\lambda}^{\alpha}$, the curvature tensor
$R_{\;\beta\mu\nu}^{\alpha}$, the Ricci tensor $R_{\mu\nu}=R_{\;\mu\alpha\nu
}^{\alpha}$ and the length curvature $F_{\mu\nu}=\partial_{\nu}\sigma_{\mu
}-\partial_{\mu}\sigma_{\nu}.$ The simplest invariant scalars, in
four-dimensional space-time, that can be constructed out of these are:
$\sqrt{-g}R^{2}$,$\sqrt{-g}R_{\alpha\beta\mu\nu}R^{\alpha\beta\mu\nu}%
,\sqrt{-g}R_{\alpha\beta}R^{\alpha\beta}$ and $\sqrt{-g}F_{\alpha\beta
}F^{\alpha\beta}$, where $R=g^{\alpha\beta}R_{\alpha\beta}$ denotes the Ricci
scalar calculated with the Weyl affine connection. (Curiously, the first of
these invariants appears in the action of some F(R) theories, for instance, in
the well known Starobinky's model of inflation \cite{Starobinsky}))

For reasons of consistency of his physics with the new geometry, Weyl required
his theory to be completely invariant with respect to change between gauges
(or frames). On the other hand, he chose the simplest of all possible
invariant actions, namely,%
\begin{equation}
S=\int d^{4}x\sqrt{\left\vert g\right\vert }[R^{2}+\omega F_{\mu\nu}F^{\mu\nu
}], \label{action}%
\end{equation}
where $\omega$ is a constant \footnote{Here we are not considering the matter
action.}. This action describes the gravitational-electromagnetic sector only.
(Incidentally, it is odd that Weyl did not consider the coupling with matter,
which clearly constitutes an element of incompleteness of the theory. We shall
return to this point later.) Carrying out variations with respect to
$\sigma_{\mu}$ and $g_{\mu\nu}$ \ will lead, respectively, to the following
field equations:
\begin{equation}
\frac{1}{\sqrt{-g}}\partial_{\nu}\left(  \sqrt{-g}F^{\mu\nu}\right)  =\frac
{3}{2\omega}g^{\mu\nu}\left(  R\sigma_{\nu}+\partial_{\nu}R\right)  ,
\label{maxwell}%
\end{equation}%
\begin{equation}
R(R_{(\mu\nu)}-\frac{1}{4}g_{\mu\nu}R)=\omega T_{\mu\nu}-D_{\mu\nu},
\label{einstein-weyl}%
\end{equation}
where \ $R_{(\mu\nu)}$ stands for the symmetric part of $R_{\mu\nu}$%
,\ $T_{\mu\nu}=F_{\mu\alpha}F_{\;\nu}^{\alpha}+\frac{1}{4}g_{\mu\nu}%
F_{\alpha\beta}F^{\alpha\beta}$ and $D_{\mu\nu}=\nabla_{(\mu}\nabla_{\nu
)}R+\frac{1}{2}R(\sigma_{\mu;\nu}+\sigma_{\nu;\mu})+R\sigma_{\mu}\sigma_{\nu
}+R_{,\mu}\sigma_{\nu}+R_{,\nu}\sigma_{\mu}$. Note that the presence of the
term $D_{\mu\nu}$ introduces derivatives of third and forth order in the
theory. This fact was readily pointed out by Pauli, who considered it to be a
flaw of Weyl theory \cite{Pauli2}. (However, as is well known, present-day
researchers welcome higher-derivative theories since, as was later shown, they
allow renormalizability of divergences in the quantum corrections to the
interactions of matter fields \cite{Steller} .)

The above equations are drastically simplified if we choose the so-callled
\textit{natural gauge}, defined by Weyl as $R=\Lambda=const\neq0.$ In this
case (\ref{maxwell}) and (\ref{einstein-weyl}) reduce to
\begin{equation}
\frac{1}{\sqrt{-g}}\partial_{\nu}\left(  \sqrt{-g}F^{\mu\nu}\right)
=\frac{3\Lambda}{2\omega}\sigma^{\mu}, \label{m1}%
\end{equation}%
\begin{equation}
\widetilde{R}_{\mu\nu}-\frac{1}{2}\widetilde{R}g_{\mu\nu}+\frac{\Lambda}%
{4}g_{\mu\nu}+\frac{3}{2}(\sigma_{\mu}\sigma_{\nu}-\frac{1}{2}g_{\mu\nu}%
\sigma^{\alpha}\sigma_{\alpha})=\frac{\omega}{\Lambda}T_{\mu\nu}, \label{m2}%
\end{equation}
where $\widetilde{R}_{\mu\nu}$ and $\widetilde{R}$ are now Riemannian and
defined with respect to the metric $g_{\mu\nu}$. At this point, let us note
that we can reobtain the field equations in a general gauge (\ref{maxwell})
and (\ref{einstein-weyl}) in an elegant and straigthforward way by using the
gauge transformations (\ref{conformal1}) and (\ref{gauge}) only, avoiding the
long and tedious calculations involved in the process of carrying out
variations in the action (\ref{action}) (see Appendix).

It is worth to mention that Weyl's theory correctly predicts the perihelion
precession of Mercury as well as the gravitation deflection of light by a
massive body \cite{Pauli2}. This is a consequence of the fact that all vacuum
solutions of Einstein's equations (including the Schwarzschild solution)
satisfy (\ref{maxwell}) and (\ref{einstein-weyl}) when we set $\sigma_{\mu}=0$.

Now before we start our discussion of the Einstein's objection to Weyl's
theory, we would like to stress that to build his theory Weyl adopted a very
strong and, at the same time, rather restrictive principle, the so-called
\textit{Principle of Gauge Invariance}, which requires all physical quantities
to be invariant under the gauge transformations (\ref{conformal1}) and
(\ref{gauge}). This principle was strictly followed by Weyl and guided him
\ to choose the action (\ref{action}). It should also be noted here that any
invariant scalar of this geometry must necessarily be formed from both the
metric $g_{\mu\nu}$ and the Weyl gauge field $\sigma_{\mu}.$ These two fields
constitute an essential and intrinsic part of the geometry and neither of them
can be neglected when we want to construct an invariant scalar, so they are,
in this sense, inseparable, and must always appear together.

\section{The general relativistic limit}

In this section, let us briefly examine how we can recover general relativity
from the Weyl field equations. First, let us assume that in a certain gauge we
have $\sigma=0$, which then means that the geometry becomes Riemannian and
$R_{(\mu\nu)}=R_{\mu\nu}=\ \widetilde{R}_{\mu\nu}$. In this case, $F_{\mu\nu}$
and $T_{\mu\nu}$ vanishes, and then from (\ref{maxwell}) we have
$R=\widetilde{R}=\Lambda=$ constant, which in turn implies $D_{\mu\nu}=0.$ Now
from (\ref{einstein-weyl}) we are left with two possibilities: $\widetilde
{R}=0$ or $\widetilde{R}_{\mu\nu}=\frac{1}{4}\Lambda g_{\mu\nu}.$ In the first
case, this means that all solutions of Einstein vacuum equations (with
vanishing cosmological constant) are included. In the second case, the Weyl
vacuum solutions correspond to spaces of constant Ricci curvature (Einstein
spaces), and this seems to make the cosmological constant appear in a natural
way, deduced directly from the field equations \footnote{Note \ that the same
results follow easily from the Weyl equations written in the natural gauge.}.
(Incidentally, if $\Lambda>0$, one may be tempted to consider this fact as an
indication that Weyl theory might naturally lead to the idea that the empty
space-time of special relativity should be identified to the de Sitter space,
a speculation which has gained more attention recently after the discovery of
the acceleration expansion of the Universe \cite{de sitter}.) However, if
$\Lambda$ is sufficiently small \ its effects in the field equations can be
neglected, and then Weyl's field equations becomes identical to the Einstein
vacuum equations and the results of the so-called solar system tests satisfied
by general relativity will be in accordance with Weyl theory. Moreover, if
$\Lambda=0$ \ then Weyl's theory includes special relativity as a particular case.

\section{The geometrized electromagnetic field}

It is certainly undeniable that the geometric structure found by Weyl in his
attempt to unify gravity and electromagnetism leading in a very natural way to
the appearance of the geometric tensor $F_{\mu\nu}$ whose algebraic and
invariant properties exhibit striking similarities with Faraday tensor.
However, looking into the field equations derived from the action
(\ref{action}) chosen by Weyl deviations from Maxwell equations become
apparent. For instance, let us consider the field equations written in the
natural gauge. The equation (\ref{m1}) tell us that the electromagnetic field
is coupled to itself, i.e. it acts as its own source. On the other hand, in
(\ref{m2}) there are non-linear terms in $\sigma$, which are more
characteristic of non-linear theories of electrodynamics. It is also
instructive to have a look at the action (\ref{action}), which, when it is
written in the natural gauge \cite{Bazin} is given by%
\[
S=\int d^{4}x\sqrt{-g}[\widetilde{R}+\frac{\omega}{2\Lambda}F_{\mu\nu}%
F^{\mu\nu}+\frac{3}{2}\sigma_{\mu}\sigma^{\mu}-\frac{\Lambda}{2}]
\]
which is equivalent to the action of the Proca's neutral spin-1 field in
curved space-time with the cosmological constant \cite{Greiner}.

Another problem of Weyl's geometrized electromagnetism concerns the motion of
\ neutral and electric charged particles. Because the affine geodesics are the
only curves which are gauge-invariant one would expect that they would
describe the motion of particles interacting only with the gravitational and
electromagnetic field. However, it is clear that from the geodesic equations
one cannot obtain the equation of motion for a charged particle moving in a
curved space-time, i.e, the Lorentz force equation. Let us recall that in
special or general relativity the Lorentz force appears when we perform
variations in the action $S=\int d^{4}x\sqrt{-g}A_{\mu}dx^{\mu}$ containing
the interaction of charged particle with the electromagnetic 4-potential
$A_{\mu}$. In Weyl theory, there is no prescription of how matter interacts
with the gravitation and the (geometric) electromagnetic field, an element of
"incompleteness" that we shall consider later, in Section 7.

\section{The problem of time in Weyl's theory}

It has been recognized in recent years that the notion of time in Weyl's
theory is rather problematic. To begin with, let us recall Einstein's
objection contained in an addendum to Weyl's original paper concerning the
dependence of the clock rate of ideal clocks on their paths \cite{Weyl}. This
is now referred to as the \textit{second clock effect} \cite{Penrose}. In
order to examine Einstein's objection, let us first make more explicit the
hypotheses upon which the argument is based, which may be stated as follows:

i) The proper time $\triangle\tau$ measured by a clock travelling along a
curve $\alpha=\alpha(\lambda)$ is given as in general relativity, namely, by
the (Riemannian) prescription
\begin{equation}
\triangle\tau=\frac{1}{c}\int\left[  g(V,V)\right]  ^{\frac{1}{2}}%
d\lambda=\frac{1}{c}\int\left[  g_{\mu\nu}V^{\mu}V^{\nu}\right]  ^{\frac{1}%
{2}}d\lambda, \label{proper time}%
\end{equation}
where $V$ denotes the vector tangent to the clock's world line and $c$ is the
speed of light. This supposition is known as the \textit{clock hypothesis }and
clearly assumes that the proper time only depends on the instantaneous speed
of the clock and \ on the metric field \cite{d'Inverno}. (Note that the gauge
field $\sigma$, which is also an essential and unseparable part of the Weyl
space-time geometry does not apper in the above expression) \cite{d'Inverno}.

ii) The clock rate of a clock (in particular, atomic clocks) is modelled by
the (Riemannian) length $L=$ $\sqrt{g(\Upsilon,\Upsilon)}$ of a certain vector
$\Upsilon$. As the clock moves in space-time $\Upsilon$ is
parallel-transported along its worldline from a point $P_{0}$ to a point $P$,
hence $L=L_{0}e^{\frac{1}{2}\int\sigma_{\alpha}dx^{\alpha}}$, $L_{0}$ and $L$
denoting the duration of the clock rate of the clock at $P_{0}$ and $P$,
respectively. (Later, this assumption was made explicit by Ehlers, Pirani and
Schild \cite{eps}.)

Let us now examine more closely these two assumptions. We start with the first
hypothesis (A). First, for consistency with the \textit{Principle of Gauge
Invariance }proper time should be a gauge-invariant concept, and clearly this
requirement is not fulfilled by (\ref{proper time}). It turns out, however,
that up to this date no such invariant notion of proper time consistent with
Weyl's theory (and which does not lead to the second clock effect) is known
\footnote{In 1986, V. Perlick proposed a new notion of proper time defined in
a Weyl manifold that is invariant by Weyl transformations \cite{Perlick} and
reduces to the WIST\ and general relativistic definitions in the appropriate
limits. However, it has been shown that Perlick's time also leads to the
second clock effect \cite{avalos}.}. Secondly, in the second hypothesis, gauge
invariance is again violated as the concept of clock rate is not modelled
as\ a gauge-invariant physical quantity, let alone the fact that the Weyl
geometrical field plays no role in its determination.

Incidentally, it should be mentioned that a new notion of proper time,
entirely consistent with the Principle of Gauge Invariance, was given by V.
Perlick \cite{Perlick}. His line of reasoning is the following. In Riemannian
geometry, the compatibility condition between the metric and the connection
may be given, as we know, by the equation
\begin{equation}
\nabla_{V}\left[  g(W,U)\right]  =g\left(  \nabla_{V}W,U\right)  +g\left(
W,\nabla_{V}U\right)  , \label{compatibility 2}%
\end{equation}
where $V,W$ and $U$ are vector fields. Now consider a curve $\alpha
=\alpha(\lambda)$ and set $V=W=$ $U=\frac{d}{d\lambda}$ , the vector tangent
to $\alpha$. Then, $\frac{d}{d\lambda}$ $g\left(  \frac{d}{d\lambda},\frac
{d}{d\lambda}\right)  =2g\left(  \nabla_{\frac{d}{d\lambda}}\frac{d}{d\lambda
}\right)  $, and we can say that $\lambda$ is the arc-length parameter $s$ of
the curve $\alpha$ (up to an affine reparametrization) \ if and only if
$g\left(  \nabla_{\frac{d}{d\lambda}}\frac{d}{d\lambda},\frac{d}{d\lambda
}\right)  =0$. \ If this condition, which may be taken to characterize the
arc-length parameter in Riemannian geometry, is carried over to Weyl geometry,
then we have a definition of proper time which is completely invariant with
respect to Weyl transformations. This was, in fact, the starting point of
Perlick's definition of proper time, which, amazingly enough, also leads to
the second clock effect \cite{avalos}. By replacing the (non-invariant)
general relativistic parametrization condition $g(\frac{d}{d\lambda},\frac
{d}{d\lambda})=1$ by the gauge-invariant equation $g\left(  \nabla_{\frac
{d}{d\lambda}}\frac{d}{d\lambda}\right)  =0$ it can be shown that the proper
time elapsed between two events corresponding to the parameter values
$\lambda_{0}$ and $\lambda$ in the curve $\alpha$\ is given by
\begin{equation}
\Delta\tau(\lambda)=\left(  \frac{d\tau/d\lambda}{\sqrt{g_{\alpha\beta}\dot
{x}^{\alpha}\dot{x}^{\beta}}}\right)  _{\lambda=\lambda_{0}}\int_{\lambda_{0}%
}^{\lambda}\exp{\left(  -\frac{1}{2}\int_{u_{0}}^{u}\sigma_{\rho}\dot{x}%
^{\rho}ds\right)  }\left[  g_{\mu\nu}\dot{x}^{\mu}\dot{x}^{\nu}\right]
^{1/2}du, \label{functional}%
\end{equation}
where dot means derivative with respect to the curve's parameter \cite{avalos}
. It has also been shown that Perlick's notion of time has all the properties
a definition of proper time in a Weyl space-time should have, such as,
Weyl-invariance, positive definiteness, additivity. Moreover, in the limit in
which the length curvature $F_{\mu\nu}$ goes to zero Perlick's time\ reduces
both to the Einsteinian proper time and to the proper time defined in Weyl
geometric scalar-tensor theories \cite{Novello}. Furthermore, it has been
proved the equivalence between Perlick's definition of proper time and the one
given in the paper by Ehlers, Pirani, and Schild (EPS) \cite{avalos,eps},
which was entirely based on an axiomatic approach
\cite{avalos,eps,Teyssandier} .

It is interesting to note that Perlick's proposal leads to a new kind of
geometry. Indeed, one can view the equation (1) as a prescription of how to
define length of curves in an entire class of Weyl manifolds. In other words,
Perlick's proper time endows space-time with new metric properties which are
distinct from the Riemannian metric of each member of the class. At this
point, one may ask the question: What are the \textquotedblleft
geodesics\textquotedblright\ of this geometry? This question is motivated both
by geometry and physics. Indeed, one may be interested in knowing whether it
is possible or not to define \textquotedblleft distance\textquotedblright%
\ between points. Or, one may regard the geodesics as describing the paths of
freely falling particles. In any case, the equation of the curve which
extremizes Perlick's functional is needed. \ The answers to these questions
have not been found yet as the extremization of the functional
(\ref{functional}) does not seem easy to carry out. However, preliminary
results yield the following: i) Perlick's geodesics do not coincide with the
affine geodesics of Weyl geometry; ii) they exhibit a non-local character in
the sense that they depend on the whole past of world line of the particle. In
fact, Perlick's geometry is completely non-local and this non-locality should
not be unexpected since the second clock effect may also be viewed as a
non-local phenomenon. In addition to the fact that Perlick's\ time introduces
great mathematical difficulties (in spite of being a gauge-invariant notion),
it is not operational when it comes to consider matter in Weyl's theory, a
question to be dealt with in the next section.

\section{Matter coupling in Weyl's theory}

Let us begin by quoting some words by the British matematician M. Atiyha
regarding Einstein's historical objection to Weyl's theory: \textit{"Given
this devastating critique it is remarkable but fortunate that Weyl's paper was
still published... Clearly the beauty of the idea attracted the editor..."}
\cite{Atiyha}\ Certainly this "devastating" critique was what prevented Weyl
from going ahead and completing his elegant and aesthetically appealing theory
by adding matter to his universe and get the full field equations in the
presence of matter. In what follows we shall briefly discuss this point and
suggest a possible way of carrying out this completion while maintaining
consistency with the principle of gauge invariance. In this article we shall
just outline the general theoretical construction, \ a preliminary step in
this direction, leaving applications for future work.

We start by calling attention of the reader to the fact that in the case of
the so-called geometric scalar-tensor gravity theories the definition of
proper time is given by the gauge-invariant equation \cite{Romero}
\begin{equation}
\Delta\tau=\int_{a}^{b}e^{-\frac{\phi}{2}}\left(  g_{\mu\nu}\frac{dx^{\mu}%
}{d\lambda}\frac{dx^{\nu}}{d\lambda}\right)  ^{\frac{1}{2}}d\lambda.
\label{wist time}%
\end{equation}
These theories are framed in a geometric structure known in the literature as
\textit{WIST} (Weyl integrable space-time) \cite{Novello}. It can be viewed as
a "weak" version of Weyl's geometry when the $1$-form $\sigma$ representing
the gauge field is exact, i.e. $\sigma=d\phi$. Thus instead of the
geometrizing the electromagnetic field we shall geometrize a scalar field. In
this case, the Weyl transformations (\ref{conformal1}) and (\ref{gauge})
become $\overline{g}=e^{f}g$ $\ $and $\ \overline{\phi}=\phi+$ $f$\ ,\ and the
compatibility condition now is given by \bigskip%
\[
\nabla_{V}\left[  g(W,U)\right]  =g\left(  \nabla_{V}W,U\right)  +g\left(
W,\nabla_{V}U\right)  +d\phi(V)g(W,U),
\]
in which $U,$ $V$ and $W$ are vector fields.The relevant point we wish to
highlight here is that (\ref{wist time}) is nothing more than the clock
hypothesis redefined in terms of the \textit{gauge-invariant }metric
$\gamma_{\mu\nu}=\ e^{-\frac{\phi}{2}}g_{\mu\nu}$. This gives us a clue to
tackle the problem of matter coupling in the non-integrable case. For this
purpose, let us recall the procedure used to define an invariant
energy-momentum tensor in geometric scalar-tensor theories. Let $S^{(m)}=\int
d^{4}x\sqrt{\left\vert \eta\right\vert }\mathcal{L}_{m}(\psi_{A},\partial
\psi_{A})$ denote the action of the matter fields $\psi_{A}$ (which we want to
couple with the geometry) in Minkowski space-time. We next apply the
\textit{principle of minimal coupling }$\eta_{\mu\nu}\rightarrow\gamma_{\mu
\nu},\partial\psi_{A}\rightarrow\nabla\psi_{A}$, where $\nabla$ stands for the
covariant derivative with respect to the metric connection determined by
$\gamma_{\mu\nu}$. We proceed to define the gauge-invariant energy-momentum
tensor $T_{\mu\nu}$ (the source of the gravitational field) by the well-known
Hilbert prescription
\begin{equation}
\delta S^{(m)}=k\int d^{4}x\sqrt{\left\vert \gamma\right\vert }T_{\mu\nu
}^{(m)}\delta\gamma^{\mu\nu}, \label{poster04}%
\end{equation}
with $k$ denoting the coupling constant. In this way we have an invariant
procedure to obtain the coupling between matter and space-time.

Our aim at this point is to obtain a gauge-invariant procedure which enable us
to construct the coupling between matter and geometry in Weyl theory by
following a somehow similar procedure as in above. Clearly, the
"non-locallity" of the functional (\ref{functional}) makes the use of
Perlick's metric virtually impossible. Therefore we need to find out another
gauge-invariant metric tensor. It turns out that Weyl's idea of working out
the field equations in a particular gauge, namely, the natural gauge defined
by the condition $R=\Lambda$ may be of great help. Indeeed, by just picking up
this idea we are now able to define a metric tensor which may be regarded as
the representative of the whole conformal structure $\mathcal{M=}\left\{
(g,\nabla,\sigma)\right\}  $ of the Weyl manifold. Indeed, let $(g,\nabla
,\sigma)$ be an arbitrary member of $\mathcal{M}$ and define the tensor
$\boldsymbol{\gamma=}\frac{R}{\Lambda}g$ for some $\Lambda>0$\ \footnote{If
$R=0$ \ it is easy to verify that Weyl field equations become trivial and we
have no longer an electromagnetic field.}. \ Now suppose that $(\overline
{g},\nabla,\overline{\sigma})$ is another member of $\mathcal{M}$. Clearly
both members are related by the transformations $\overline{g}_{\mu\nu}=$
$e^{f}g_{\mu\nu}$ , $\overline{\sigma}_{\alpha}=\sigma_{\alpha}+$
$\partial_{\alpha}f$, \ for some function $f$\ . \ Thus, the fact that
$\overline{R}=\overline{g}^{\mu\nu}\overline{R}_{\mu\nu}=\overline{g}^{\mu\nu
}R_{\mu\nu}=e^{-f}g^{\mu\nu}R_{\mu\nu}=e^{-f}R$ immediately implies
$\overline{\boldsymbol{\gamma}}=\frac{\overline{R}}{\Lambda}\overline{g}%
=\frac{R}{\Lambda}g=\boldsymbol{\gamma\ },\boldsymbol{\ }$and this means that
$\boldsymbol{\ \gamma\ \ \ }$is a gauge-invariant object, which can be
computed from any member of the conformal structure $\mathcal{M}$. Note that
in Weyl natural gauge (determined by the chosen $\Lambda)$ the metric tensor
$\boldsymbol{\gamma\ \ }$assumes its simplest form, namely,
$\boldsymbol{\gamma\ }=g.$ The same reasoning lead us to define a second
gauge-invariant object, namely, the $1$-form given by $\boldsymbol{\xi}%
$\ $=\sigma+d(\ln R)$. Therefore, we can regard $\boldsymbol{\xi\ }\ $as the
$1$-form representative of $\mathcal{M}$. Once we have the gauge-invariant
metric tensor $\boldsymbol{\gamma\ }$\ we can then adopt the same procedure
used in WIST theories to obtain a gauge invariant energy-momentum tensor in
Weyl theory by defining $\delta S^{(m)}=\kappa\int d^{4}x\sqrt{\left\vert
\boldsymbol{\gamma}\right\vert }T_{\mu\nu}^{(m)}\delta\boldsymbol{\gamma}%
^{\mu\nu}$. Therefore, our strategy will be to reframe Weyl's theory in terms
of $\boldsymbol{\gamma\ }$\ and $\boldsymbol{\xi\ }$. \ \ \ \ \ \ 

\section{A new approach to Weyl's theory}

In this section we summarize the ideas devoloped so far in the following set
of postulates:

P1. We still consider space-time modelled by the conformal structure
$\mathcal{M}$, whose members are related by the group of transformations
(\ref{conformal1}) and (\ref{gauge}). However, all relevant geometric objects
will be constructed from \ $\boldsymbol{\gamma\ \ }$and $\boldsymbol{\xi\ .}$

P2. The field equations of the theory will be given by varying the action
$S=\sqrt{\left\vert \boldsymbol{\gamma\ }\right\vert }[R^{2}+\omega F_{\mu\nu
}F^{\mu\nu}+\mathcal{\varkappa L}_{m}]d^{4}x$ with respect to
$\boldsymbol{\gamma\ }$, $\boldsymbol{\xi}$ and $\psi_{A}$, where
$\mathcal{\varkappa}$ is a coupling constant and $\mathcal{L}_{m}(\psi
_{A},\nabla\psi_{A})$ denote the Lagrangian of the matter fields. In the Weyl
gauge, these variations will have the form%

\begin{equation}
\delta S=\delta\int d^{4}x\sqrt{-g}[R+\frac{\omega}{2\Lambda}F_{\mu\nu}%
F^{\mu\nu}-\frac{\Lambda}{2}+\kappa\mathcal{L}_{m}]=0,\label{Weyl new 1}%
\end{equation}
yielding the equations
\begin{equation}
\widetilde{R}_{\mu\nu}-\frac{1}{2}\widetilde{R}g_{\mu\nu}+\frac{\Lambda}%
{4}g_{\mu\nu}+\frac{3}{2}(\sigma_{\mu}\sigma_{\nu}-\frac{1}{2}g_{\mu\nu}%
\sigma^{\alpha}\sigma_{\alpha})=\frac{\omega}{\Lambda}T_{\mu\nu}-\kappa
T_{\mu\nu}^{(m)}\label{m 1a}%
\end{equation}%
\begin{equation}
\frac{1}{\sqrt{-g}}\partial_{\nu}\left(  \sqrt{-g}F^{\mu\nu}\right)
=\frac{3\Lambda}{2\omega}\sigma^{\mu},\label{m 2}%
\end{equation}%
\[
\Phi^{A}=0
\]
where $\delta\int d^{4}x\sqrt{-g}[\kappa\mathcal{L}_{m}]=\int d^{4}x\sqrt
{-g}\Phi^{A}\delta\psi_{A}$ , and $\kappa=\frac{\mathcal{\varkappa}}{2\Lambda
}$.

Finally, to complete the theoretical framework we assume the following postulates:

P3. The motion of free-falling test particles \ will be given by Riemannian
geodesics with respect to the metric tensor $\boldsymbol{\gamma\ .}$

P4. The \ gauge-invariant proper time of a standard clock will be given by
assuming the usual clock hypothesis, namely, that%
\[
\triangle\tau=\frac{1}{c}\int\left[  \gamma(V,V)\right]  ^{\frac{1}{2}%
}d\lambda,
\]
which in Weyl gauge reduces to (\ref{proper time}).

P5.The clock rate of standard clocks are strictly determined by the metric
properties of $\boldsymbol{\gamma}$ and its corresponding Levi-Civita connection.

As an application of the ideas developed so far, we shall now solve the field
equations (\ref{Weyl new 1}) and (\ref{m 1a}) assuming a simple cosmological scenario.

\section{A simple cosmological solution}

Although \ general relativity is still considered the best available theory of
gravity, and as such has been applied to the study of the universe with
enormous success, we are currently seeing a great interest (which can be
justified for several reasons) in alternative theoretical proposals, generally
referred to as "modified theories of gravity". This kind of research is
particularly connected and stimulated by the recent advances in the field of
observational cosmology \cite{Ferreira}. We thus thought it could be
interesting to apply the \ present new approach to Weyl's theory in searching
a solution of the field equations in a simple cosmological setting.

The model assumes homogeneity and isotropy both in the metric and the vector
field, the first being assumed to be given by a Friedmann-Robertson-Walker
line element, in which, for simplicity, we have chosen a flat spatial section
($k=0$). As to the matter distribution of the universe, we admit that it is
described by the energy-momentum tensor of a perfect fluid $T_{\mu\nu}%
^{(m)}=(\rho+p)u_{\mu}u_{\nu}-pg_{\mu\nu}$, where $\rho$, $p$, and $u_{\mu}$
denotes the energy-density, the pressure and the $4$-velocity of the fluid.
(We do not assume a particular equation of state, leaving it to be determined
by the solution of the field equations.)

We thus write $ds^{2}=dt^{2}-a^{2}(t)(dx^{2}+dy^{2}+dz^{2})$ \ for the line
element, and $\sigma_{\mu}=(\phi(t),\theta(t),\theta(t),\theta(t))$ for the
Weyl vector field. A direct calculation give us $F^{\mu\nu}=\frac{\dot{\theta
}}{a^{2}}$ $\left(
\begin{array}
[c]{cccc}%
0 & 1 & 1 & 1\\
-1 & 0 & 0 & 0\\
-1 & 0 & 0 & 0\\
-1 & 0 & 0 & 0
\end{array}
\right)  $ and $T_{\mu\nu}=\frac{\dot{\theta}^{2}}{2}$ $\left(
\begin{array}
[c]{cccc}%
\frac{3}{a^{2}} & 0 & 0 & 0\\
0 & 1 & -2 & -2\\
0 & -2 & 1 & -2\\
0 & -2 & -2 & 1
\end{array}
\right)  ,$where dot denotes derivative with respect to $t.$

Let us now work with the equations in the natural gauge. Setting $\mu=0$ in
(\ref{m 2}) will give us%
\begin{equation}
\partial_{0}(a^{3}F^{0\nu})=\frac{3\Lambda}{2\omega}a^{3}\sigma^{0}=0,
\end{equation}
which then implies $\phi(t)=0.$ On the other hand, setting $\mu=1$ in the same
equation leads us to
\begin{equation}
\overset{..}{\theta}+(\frac{\dot{a}}{a})\dot{\theta}=\frac{3\Lambda}{2\omega
}\theta.\label{cosmo 1}%
\end{equation}
Now let us consider the equation (\ref{m 1a}). For $\mu=1$ and $\nu=2$ we
obtain
\[
\frac{\dot{\theta}}{\theta}=\pm\sqrt{-\frac{3\Lambda}{2\omega}},
\]
which immediately yields the solutions
\[
\theta=e^{\pm\sqrt{-\frac{3\Lambda}{2\omega}}t},
\]
in which the integration constant was set equal to unity by rescaling the line
element \footnote{Because we would like to interpret $\Lambda$ as the
cosmological constant we are restricting ourselves to negative values of
$\omega.$}. Choosing the negative solution above and inserting it in
(\ref{cosmo 1}) gives for the scale factor $a(t)$ \footnote{Choosing the
positive sign here leads to a rather non-phyical scenario, which merely
describes a contracting universe.}
\[
a(t)=e^{2\sqrt{-\frac{3\Lambda}{2\omega}}t}\text{ ,}%
\]
in which the integration constants were set equal to unity just by rescaling
the line element and choosing \ appropriate initial conditions. Setting
$\mu=\nu=0$ in we obtain
\[
\rho=-\frac{\Lambda}{\kappa}(\frac{18}{\omega}+\frac{1}{4})-\frac{9}{2\kappa
}e^{^{-6\sqrt{-\frac{3\Lambda}{2\omega}}t}},
\]
whereas putting $\mu=\nu=1$ leads to
\[
p=\frac{\Lambda}{\kappa}(\frac{18}{\omega}+\frac{1}{4})
\]

The solution obtained above represents a typical non-singular and expansive
model, that is, a de Sitter universe. Solutions of this kind have been found
in different contexts suggesting the possibilitity of describing dark energy
in our present universe or inflation in the early universe. It is interesting
to note that $\sigma(t)\rightarrow0$ when $t\rightarrow\infty$. In other
words, the Weyl field tends to fade away with the expansion of the universe,
while the equation of state of the cosmological fluid becomes $p=-\rho$, which
is typical of dark energy models.\ It should be mentioned that inhomogenous
time-dependent equations have also been considered in dark some
energy\ scenarios \cite{Odintsov}. Finally, in connection with the simple
model outlined above, we would like to call attention of the reader to the
fact that, although most versions of inflationary cosmology require a scalar
field, previous results found in the literature show that inflation can also
be driven by vector fields, including the particular case of massive fields
\cite{Ford}.

\section{Final remarks}

We would like to conclude this work with a few remarks. First of all, it is
important to stress the fact that the adoption of \ a very special set of
gauge-invariant tensors playing the role of representatives of the space-time
modelled as a conformal structure leads to two unexpected consequences: i)
non-local effects, such as the second clock effect are no longer predicted;
ii) the coupling between space-time and matter is carried out in an invariant
way following the traditional prescription contained in the principle of
minimal coupling of general relativity.

In deriving the equation for the Weyl field $\sigma$ when matter fields are
present we have implicitly made the assumption that $\mathcal{L}_{m}$ does not
depend on $\sigma$, or in other words, that the vector field does not couple
directly with matter. Surely, if one wishes a more general framework, it is
possible to weaken this restriction by just adding a current term $j_{\mu}$,
given by $\delta\int d^{4}x\sqrt{-g}[\kappa\mathcal{L}_{m}]=$ $\int
d^{4}x\sqrt{-g}j_{\mu}\delta\sigma^{\mu}$.

Finally, the original identification of the 1-form field $\sigma$ with the
eletromagnetic potential is no longer assumed here. Instead, the pair
$(\boldsymbol{\gamma\ ,\xi\ })$ constitutes what would we would call the
complete gravitational field. In this way, we are simply left with a modified
gravity theory instead of a unified theory as in the Weyl's original program.

\section*{Acknowledgements}

\noindent The authors would like to thank R. Avalos for helpful discussions.
This work was supported by CAPES and CNPq.

\section{Appendix}

In what follows we show how to obtain the Weyl field equations (\ref{maxwell})
and (\ref{einstein-weyl}), written in an arbitrary gauge, directly from the
equations (\ref{m1}) and (\ref{m2}), the latter valid in the natural gauge only.

We start by looking for a Weyl transformation (\ref{conformal1}) and
(\ref{gauge}) which allows us to go from an arbitrary gauge $(M,\overline
{g},\overline{\sigma})$ to a particular gauge $(M,g,\sigma)$, in which
$R=\Lambda\neq0$, where $\Lambda$ is an arbitrary constant. From the fact that
the Ricci tensor is gauge-invariant it is not difficult to verify that the
desired transformation is given by taking $f=\ln($ $\frac{\overline{R}%
}{\Lambda})$, where $\overline{R}$ denotes the Ricci scalar in the arbitrary
gauge $(M,\overline{g},\overline{\sigma})$. We thus have $g_{\mu\nu}%
=\frac{\overline{R}}{\Lambda}\overline{g}_{\mu\nu}$ and $\sigma_{\mu
}=\overline{\sigma}_{\mu}+\frac{1}{\overline{R}}\partial_{\mu}\overline{R}$.
Let us now rewrite the equation (\ref{m2}) in terms of $R_{\mu\nu}$ and $R$
(respectively, the Ricci and scalar curvature calculated with the Weyl
connection), also recalling the identity which relates the two Ricci tensors
$R_{\mu\nu}$ and $\widetilde{R}_{\mu\nu}$, the first calculated with the Weyl
connection and the second with the Christoffel symbols:
\begin{equation}
\widetilde{R}_{\mu\nu}=R_{(\mu\nu)}+\frac{1}{2}(\widetilde{\nabla}_{\mu}%
\sigma_{\nu}+\widetilde{\nabla}_{\nu}\sigma_{\mu}+g_{\mu\nu}\widetilde{\nabla
}_{\alpha}\sigma^{\alpha})+\frac{1}{2}(\sigma_{\mu}\sigma_{\nu}-g_{\mu\nu
}\sigma_{\alpha}\sigma^{\alpha}), \label{id 1}%
\end{equation}
the symbol $\widetilde{\nabla}$ \ standing for the Riemannian covariant
derivative. Contracting the above equation with $g^{\mu\nu}$ yields
\begin{equation}
\widetilde{R}=R-\frac{3}{2}\sigma_{\alpha}\sigma^{\alpha}, \label{id 2}%
\end{equation}
where we have used the fact that (\ref{m1}) implies $\widetilde{\nabla
}_{\alpha}\sigma^{\alpha}=0.$ Substituting (\ref{id 1}) and (\ref{id 2}) into
(\ref{m2}) leads to
\begin{equation}
R_{(\mu\nu)}-\frac{1}{2}g_{\mu\nu}R+\frac{\Lambda}{4}g_{\mu\nu}+\frac{1}%
{2}(\nabla_{\mu}\sigma_{\nu}+\nabla_{\nu}\sigma_{\mu})+\sigma_{\mu}\sigma
_{\nu}=\frac{\omega}{\Lambda}T_{\mu\nu}\text{ ,} \label{weyl terms}%
\end{equation}
in which we have taken into account the following relation between the Weylian
and Riemannian\ covariant derivatives:
\[
\nabla_{\mu}\sigma_{\nu}=\widetilde{\nabla}_{\mu}\sigma_{\nu}+\sigma_{\mu
}\sigma_{\nu}-\frac{1}{2}g_{\mu\nu}\sigma_{\alpha}\sigma^{\alpha}.
\]
Recalling the expression of $T_{\mu\nu}$ in Section 3 we now see that all
terms in (\ref{weyl terms}) possess a well defined transformation under
(\ref{conformal1}) and (\ref{gauge}). Therefore, chosing the latter as being
given by $g_{\mu\nu}=\frac{\overline{R}}{\Lambda}\overline{g}_{\mu\nu}$ and
$\sigma_{\mu}=\overline{\sigma}_{\mu}+\frac{1}{\overline{R}}\partial_{\mu
}\overline{R}$ we get, after a straightforward calculation,
\begin{equation}
\overline{R}_{(\mu\nu)}-\frac{1}{4}\overline{g}_{\mu\nu}\overline{R}%
+\nabla_{(\mu}\overline{\sigma}_{\nu)}+\frac{1}{\overline{R}}\nabla_{(\mu
}\nabla_{\nu)}\overline{R}+\overline{\sigma}_{\mu}\overline{\sigma}_{\nu
}+\frac{1}{\overline{R}}\overline{\sigma}_{(\mu}\nabla_{\nu)}\overline
{R}=\frac{\omega}{\overline{R}}\overline{T}_{\mu\nu}. \label{weyl 2}%
\end{equation}
Multiplying the above equation by $\overline{R}$ clearly leads to
(\ref{einstein-weyl}). Finally, the equation (\ref{maxwell}) is directly
obtained by the same procedure.

\end{document}